\documentclass[pre,showpacs,10pt,amsmath,amssymb]{revtex4}
\usepackage{graphicx}

\begin{document}

\title{Non-Markovian L\'evy diffusion in nonhomogeneous media}

\author
{T. Srokowski}

\affiliation{
 Institute of Nuclear Physics, Polish Academy of Sciences, PL -- 31-342
Krak\'ow,
Poland }

\date{\today}

\begin{abstract}

We study the diffusion equation with a position-dependent, power-law 
diffusion coefficient. The equation possesses the Riesz-Weyl fractional operator 
and includes a memory kernel. It is solved in the diffusion limit of small 
wave numbers. Two kernels are considered in detail: the exponential kernel, 
for which the problem resolves itself to the telegrapher's equation, 
and the power-law one. The resulting distributions have the form of the L\'evy 
process for any kernel. The renormalized fractional moment is introduced 
to compare different cases with respect to the diffusion properties of the system.

\end{abstract}


\maketitle

\section{Introduction}

Diffusion processes are usually described in terms of either
differential or fractional equations which
contain a constant diffusion coefficient. In many physical problems,
however, that coefficient depends on the position variable and such
dependence is important \cite{lop}. As a typical example can serve the
transport in porous, inhomogeneous media and in plasmas.  
Modelling the aggregation of interacting particles must take into
account nonlocal effects, since the particle mobility depends on the
average density \cite{hol}: the coalescence of particles results from
long-range interactions (the Poisson-Smoluchowski paradigm) and the
corresponding evolution equations contain a position-dependent
coefficient. That modelling can be accomplished directly, via the non-local
Fokker-Planck equation, in which the term with the space derivative is
multiplied by a kernel and integrated over the position \cite{mal}.
A similar method, applicable to the L\'evy processes, consists in the 
integrating over the L\'evy index, with some kernel (the distributed
order space fractional equation) \cite{che}.
The spatial inhomogeneity can be also taken into account 
as an external potential which may substantially change
the diffusive properties of the stochastic system, in particular 
of the L\'evy flights \cite{bro}.

The L\'evy distributions constitute the most general class of stable processes 
and the Gaussian distribution is their special case. One can expect that the 
L\'evy (and non-Gaussian) distributions emerge in transport 
processes for which the observable 
values experience long jumps, e.g. due to the existence of long range 
correlations. The theory of the L\'evy flights is 
applicable to problems from various branches of science and technology. 
Moreover, the handling of specific and realistic systems 
often requires taking into account both memory effects 
and inhomogeneity of the media. As a typical example of the nonhomogeneous 
problem can serve the diffusion in the porous media; 
they often display the fractal structure and the diffusion on the macro- and 
mesoscale can be expressed by a stochastic equation driven by the L\'evy 
process \cite{par}. In general, the transport in fractal media
can be described by the fractional Fokker-Planck equation with a
variable, position dependent, diffusion coefficient
\cite{met3,met4,tar}. The L\'evy flights bring about the accelerated diffusion 
in the reaction-diffusion systems \cite{cas} and the probability distribution 
for that process is expressed by the fractional Fisher-Kolmogorov equation. 
The L\'evy processes are typical for problems of high complexity, 
in particular in biological systems \cite{wes} where the fractal structures 
are also encountered. For example, the lipid diffusion in biomembranes has the 
characteristics of the L\'evy process but it can no longer be regarded as 
Markovian. The theory of Nonnenmacher \cite{non} takes into account the memory 
effects, as well as the fractal structure of the medium; 
the diffusion coefficient depends on the variable diameter of 
the holes in the solvent through which the molecules jump. 
The application of the L\'evy processes is natural also in many social 
and environmental problems. Recently, it has been established \cite{bro1} 
that the people mobility, estimated by the bank notes circulation and 
studied in terms of the stochastic fractional equations,  
strongly depends on geographical and sociological conditions. Therefore, its
study requires including position-dependent quantities. 
That problem is directly related to the spread of infectious diseases. 
It has been demonstrated in the example of the SARS epidemic and 
by means of percolation model simulations \cite{fuji} that 
the disease can spread very rapidly. Usually one assumes that the 
infection probability at a given distance is L\'evy distributed due 
to the long-range interactions but the process is local in time \cite{mol}. 
On the other hand, the percolation model of the epidemics developed 
in Ref.\cite{jim} is restricted to short-range interactions (is local in space) 
but it introduces the incubation times which obey the L\'evy statistics 
and then the model is non-Markovian.

In Refs.\cite{kam1,sro1}, the master equation for a
jumping process, stationary and Markovian, has been studied. That
process is a version of the coupled continuous time random walk (CTRW), 
defined in terms of two probability distributions: 
the Poissonian waiting time distribution with the position-dependent jumping
frequency and a jump-size distribution. The standard technique to handle 
such master equations is the Kramers-Moyal expansion which produces the
Fokker-Planck equation for the Gaussian jumping size distribution and it
yields correct results for large times and large distances \cite{uwa1}.
For the L\'evy distributed jumping size, the Fokker-Planck equation
becomes the fractional diffusion equation, with 
the Riesz-Weyl fractional operator and the variable coefficient $D(x)$.
Formally, it can be derived from the master equation by taking the Fourier 
transform and by neglecting the higher terms in the wave number expansion 
of the jumping-size distribution (the diffusion approximation) \cite{sro}. 
The equation reads 
\begin{equation}
\label{frace}
\frac{\partial p(x,t)}{\partial t}=K^\mu\frac{\partial^\mu[D(x)p(x,t)]}{\partial|x|^\mu},
\end{equation}
where $1<\mu<2$. Since the diffusion coefficient is just the jumping 
frequency, the medium inhomogeneity enters the problem via the $x$-dependent 
waiting time distribution. For the Gaussian case ($\mu=2$), all kinds 
of diffusion, both normal and anomalous, are 
predicted \cite{sro} and they are determined by the jumping frequency. 

The Eq.(\ref{frace}) can be regarded as a
special case of a more general problem than the random walk
and which traces back to the microscopic
foundations of the nonequilibrium statistical mechanics.
The well-known achievement of Zwanzig \cite{zwa} was the derivation of
the non-Markovian kinetic equation for the probability distribution in the space 
of macroscopic state variables. More precisely, by
starting from the Liouville-von Neumann equation for the density operator $\rho$:
$i\hbar\partial \rho/\partial t=[H,\rho]$, where $H$ is the Hamiltonian
of the system, one can obtain the generalized master equation:
\begin{equation}
\label{zwan}
\frac{\partial P_\xi(t)}{\partial t}=\int_0^t dt'\phi(t-t')\sum_\mu[F_{\xi\mu}P_\mu(t')-F_{\mu\xi}P_\xi(t')],
\end{equation}
where $P_\xi$ denotes the diagonal elements of the density matrix and 
$F_{\mu\xi}$ are the transition rates \cite{ken}. Then the equation is non-Markovian
and it contains the memory kernel $\phi(t)$. Markovian
equations like (\ref{frace}) follow from the
generalized master equation if the memory effects are
negligible. However, usually this is not the case. 
We have already discussed the examples of the L\'evy processes, 
with power-law tails of the distribution, which exhibit memory effects. 
In fact, the importance of these effects was realized a long time ago, 
e.g. in the description of the resonance
transfer of the excitation energy between molecules \cite{ken}. The detailed
calculation for the anthracene molecules shows that the memory kernel is
exponential and the generalized, nonlocal in time, master equation must
be applied to get proper results for small times. One can
expect that memory effects are still more pronounced 
for systems with the characteristic decay rate slower than exponential,
that often happens for atomic and molecular systems
\cite{bud}. Stochastic dynamical processes are generally nonlocal in time
due to the finite time of the interaction with the environment. 
Moreover, for a stochastic system which is coupled
to a fractal heat bath via a random matrix interaction \cite{lut}, the finite
correlations emerge and its relaxation has to be described in terms of the
generalized, non-Markovian Langevin equation, with the memory friction 
term \cite{kubo,luc}. Also the speed of transport is affected by the memory.
In the non-Markovian CTRW processes it 
hampers the dynamics and such systems are subdiffusive \cite{met}. 
Such processes are described by the generalized master equation, with a memory 
kernel, if the waiting time distribution possesses long, algebraic tails. 
That equation follows directly from the generalized Chapman-Kolmogorov 
equation which determines the probability distribution 
in the phase space \cite{metk,metz}.

By applying the nearest-neighbour approximation on the transition rates $F_{\mu\xi}$
and taking the continuum limit \cite{ken}, one obtains from the Eq.(\ref{zwan})   
the non-Markovian Fokker-Planck equation. In the presence 
of long-range correlations, however, the nearest-neighbour approximation 
is no longer valid. If the transition rates are symmetric and distributed 
according to the L\'evy statistics in the continuum limit, 
the Kramers-Moyal expansion produces the fractional derivative, instead 
of the Gaussian. Then, for the variable diffusion coefficient $D(x)$, 
the equation which corresponds to the Eq.(\ref{zwan}) becomes
\begin{equation}
\label{nme}
\frac{\partial p_\gamma(x,t)}{\partial t}=\int_0^tK_\gamma(t-t'){\mathcal L}_x[p_\gamma(x,t')]dt',
\end{equation}
where the operator
\begin{equation}
\label{opl}
{\mathcal L}_x=K^\mu\frac{\partial^\mu}{\partial|x|^\mu}D(x)
\end{equation}
acts only on the $x$ variable. The parameter $\gamma$ measures the rate of 
the memory loss. 

The Eq.(\ref{nme}) is of interest both from quantum and classical point of view. 
In the atomic and molecular physics e.g. 
in a few-modes spin boson model \cite{won} and 
the random-matrix theory \cite{won1}, where the decay rate is slow, 
an equation analogous to the Eq.(\ref{nme}) 
can be applied. The operator ${\mathcal L}_x$ is then expressed in terms of
the "superoperator" which represents an instantaneous intervention of the 
environment over the system \cite{bud} and it can assume a quite general form. 
In the classical context, the Eq.(\ref{nme}) has been discussed in Ref.\cite{sok}; 
the operator ${\mathcal L}_x$ has the Fokker-Planck form in this case, with 
the constant diffusion coefficient and a potential force. 

In this paper we study the diffusion problem for systems which are driven by 
the L\'evy distributed transition rate and for which both the medium inhomogeneity and 
the memory effects are important.
We assume $D(x)=|x|^{-\theta}$ $(\theta>-1)$. 
The power-law form of the diffusion coefficient has been used to describe some
physical phenomena, e.g. the transport of fast electrons in a hot plasma \cite{ved}
and the turbulent two-particle diffusion \cite{fuj}. It is also used
in theoretical analyses of the fractional Fokker-Planck equation
\cite{len,len1,kwo1,ass}, e.g. as an ansatz 
for the problem of diffusion in the fractal media
\cite{osh,met3,met4,tar}.
Obviously, for the Markovian case $K_\gamma(t)=\delta(t)$, 
the Eq.(\ref{nme}) resolves itself to the Eq.(\ref{frace}).

In Sec.II we solve the fractional telegrapher's equation which follows from the 
Eq.(\ref{nme}) for the case of the exponential memory kernel $K_\gamma(t)$. 
The solution for an arbitrary kernel, expressed in the form of the Laplace 
transform, is derived in Sec.III. Moreover, the case of the power-law kernel 
is solved in details. In Sec.IV we derive the fractional moments and discuss 
their application to the description of the diffusion process. The results 
presented in the paper are summarized in Sec.V.

\section{The exponential kernel}

If the memory effects are weak, we can assume that the kernel $K_\gamma(t)$
decays exponentially. Then let us consider the following kernel:
\begin{equation}
\label{expk}
K_\gamma(t)=\gamma e^{-\gamma t}~~~~~~~(\gamma>0)
\end{equation}
which becomes the delta function in the limit $\gamma\to\infty$ (the Markovian case). 
In this case, the integral equation, Eq.(\ref{nme}), reduces itself
to a differential equation. It can be derived by inserting 
the Eq.(\ref{expk}) to the Eq.(\ref{nme}) and by differentiating twice over time, 
in order to get rid of the integral. Finally, we get the following equation
\begin{equation}
\label{tele}
\frac{\partial^2 p_\gamma(x,t)}{\partial t^2}+\gamma\frac{\partial p_\gamma(x,t)}{\partial t}=
K^\mu\gamma\frac{\partial^\mu[|x|^{-\theta}p_\gamma(x,t)]}{\partial|x|^\mu},
\end{equation}
which is a generalized and fractional version of the well-known 
telegrapher's equation. Originaly, the telegrapher's equation, 
which is the hyperbolic one, has been
introduced into the theory of the stochastic processes by Cattaneo
\cite{catt} in order to avoid infinitely fast propagation for very small times. 
Its fractional generalization describes e.g. a two-state process with 
the correlated noise \cite{metn} and it predicts the inhanced diffusion 
in the limit of long time. On the other hand, in the case of the divergent 
second moment, the telegrapher's equation
with the Riesz-Weyl derivative results from the fractional Klein-Kramers
equation for the L\'evy distributed jumping size \cite{metz}. In that
equation, the parameter $\gamma$ has a sense of the damping constant in the
corresponding Langevin equation. 

In the diffusion limit of small wave numbers, 
the Markovian equation, Eq.(\ref{frace}), is satisfied by 
the Fox function $H_{2,2}^{1,1}$ \cite{sro}. 
Since our main objective is to study the diffusion problem, we restrict 
also the present analysis to that limit. We will try to find the solution of 
the Eq.(\ref{tele}) in the same form as for the Markovian equation. 
Therefore we assume:
\begin{eqnarray} 
\label{s1}
p_\gamma(x,t)=NaH_{2,2}^{1,1}\left[a |x|\left|\begin{array}{c}
(a_1,A_1),(a_2,A_2)\\
\\
(b_1,B_1),(b_2,B_2)
\end{array}\right.\right],
\end{eqnarray}
where the time dependence is restricted to the function $a(t)$ and 
$N$ is the normalization constant. The method of solution, 
described in Ref.\cite{sro}, consists in 
the inserting of the Fourier transform of the expression (\ref{s1}) 
into the Fourier transformed Eq.(\ref{tele}). Then we determine the coefficients 
of the Fox function by demanding that the Eq.(\ref{tele}) should be satisfied 
in the diffusion limit, i.e. for small wave numbers. In fact, the latest 
assumption does not introduce any additional idealization since the equation 
itself is valid only in the diffusion limit. 

We start with the Fourier transform of the Eq.(\ref{tele}); it reads 
\begin{equation}
\label{fracek}
\frac{\partial^2{\widetilde p}_\gamma(k,t)}{\partial t^2}+
\gamma\frac{\partial{\widetilde p}_\gamma(k,t)}{\partial t}=
-K^\mu\gamma |k|^\mu{\mathcal F}[|x|^{-\theta}p_\gamma(x,t)].
\end{equation}
The Fourier transform of the Fox function can be expressed also in terms of 
the Fox function (for the definition and some useful properties of 
the Fox functions see Ref.\cite{sro} and references therein).
Due to the multiplication rule, the product $|x|^{-\theta}p_\gamma(x,t)$ 
is the Fox function as well. Both sides of 
the Eq.(\ref{fracek}) can now be expanded in series of fractional powers of $|k|$. 
We can satisfy the Eq.(\ref{fracek}) by a suitable choice of the parameters 
of the function (\ref{s1}) and by neglecting the terms higher than 
$|k|^\mu$. The results are the following. The expansion of 
the functions on the lhs and rhs, respectively, reads: 
${\widetilde p_\gamma}(k,t)\approx 1-N h_\mu a^{-\mu}|k|^\mu$
and ${\mathcal F}[|x|^{-\theta}p_\gamma(x,t)]\approx N h_0^{(\theta)}a^\theta$, 
with the following coefficients: $h_0^{(\theta)}=2(\mu+\theta)/(2+\theta)$ and 
$h_\mu=-2\frac{(\mu+\theta)^2}{\pi}\Gamma(-\mu)\Gamma(\mu+\theta)\cos(\mu\pi/2)
\sin(\frac{\mu+\theta}{2+\theta}\pi)$. The vanishing of all other terms of 
the order less than $\mu$ is the necessary condition to satisfy the 
Eq.(\ref{fracek}). The solution takes the form
\begin{eqnarray} 
\label{solp}
p_\gamma(x,t)=NaH_{2,2}^{1,1}\left[a|x|\left|\begin{array}{l}
(1-\frac{1-\theta}{\mu+\theta},\frac{1}{\mu+\theta}),(1-\frac{1-\theta}{2+\theta},\frac{1}{2+\theta})\\
\\
(\theta,1),(1-\frac{1-\theta}{2+\theta},\frac{1}{2+\theta})
\end{array}\right.\right]
  \end{eqnarray}
and the coefficients $b_1$ and $B_1$ are responsible for the distribution 
behaviour near $x=0$. $b_1$ and $B_1$ cannot be determined in the diffusive limit
and they are meaningless from the point of view of the diffusion 
process; the values 
$\theta$ and $1$ we have assumed correspond to the jumping process, 
considered in Ref.\cite{sro}. Generally, the Eq.(\ref{nme}) is satisfied by 
the function (\ref{solp}) for any choice of the coefficients $b_1$ and $B_1>0$,
such that $b_1\to0$ and $B_1\to1$ for $\theta\to0$.
The normalization factor $N$ can be determined in a simple way from 
the formula $N=[2\chi(-1)]^{-1}$, where $\chi(-s)$ is the Mellin transform 
of the Fox function. A simple algebra yields
\begin{equation} 
\label{nor}
N=-\frac{\pi}{2}\left[\Gamma(1+\theta)\Gamma\left(-\frac{\theta}{\mu+\theta}
\right)\sin\left(\frac{\theta}{2+\theta}\pi\right)\right]^{-1}.
\end{equation}

Alternatively, since $|k|$ is small, we can express the Fourier transform 
of the solution as 
\begin{equation}
\label{pfou}
{\widetilde p_\gamma}(k,t)\approx 1-\sigma^\mu|k|^\mu \approx \exp(-\sigma^\mu|k|^\mu), 
\end{equation}
where 
\begin{equation} 
\label{sigmu}
\sigma^\mu=K^{-\mu}\frac{(\mu+\theta)^2\Gamma(-\mu)\Gamma(\mu+\theta)
\cos(\mu\pi/2)\sin(\frac{\mu+\theta}{2+\theta})}{\Gamma(1+\theta)
\Gamma(-\frac{\theta}{\mu+\theta})\sin(\frac{\theta}{2+\theta}\pi)}\,a^{-\mu}.
\end{equation}
The Eq.(\ref{pfou}) means that the solution of Eq.(\ref{nme}) coincides with the L\'evy 
process in the limit $k\to 0$. 
Then the solution (\ref{solp}) can be expressed in the form which is generic
for any symmetric L\'evy distribution \cite{sch}:
\begin{eqnarray} 
\label{solp0}
p_\gamma(x,t)=\frac{1}{\mu\sigma}H_{2,2}^{1,1}\left[\frac{|x|}{\sigma}\left|\begin{array}{l}
(1-1/\mu,1/\mu),(1/2,1/2)\\
\\
(0,1),(1/2,1/2)
\end{array}\right.\right].
\end{eqnarray}
The formula (\ref{sigmu}) establishes the relation between the solutions 
(\ref{solp}) and (\ref{solp0}). Those expressions are equivalent only in the limit 
$k\to 0$ and they behave differently for small $|x|$. 
We will demonstrate in Sec.III that the L\'evy process is the solution of 
the Eq.(\ref{nme}) for any kernel. Therefore, the form (\ref{solp0}) 
is quite universal and we apply it in the following. The problem is reduced 
in this way to evaluating the function $\sigma(t)$. 

Now, the Eq.(\ref{fracek}) becomes the ordinary differential equation:
\begin{equation}
\label{telt}
\frac{1}{\gamma}\ddot\xi=-\dot\xi+K^\mu\frac{h_0^{(\theta)}}{h_\mu}\xi^{-\theta/\mu},
\end{equation}
where $\xi(t)=a^{-\mu}$. We assume the following initial conditions: 
$\xi(0)=\dot\xi(0)=0$ which correspond to the condition $p_\gamma(x,0)=\delta(x)$. 
The Eq.(\ref{telt}) has the structure of the equation of motion with a "friction 
term", a positive "driving force", and a "mass" $1/\gamma$. The meaning of the quantity 
$\xi$, the time evolution of which the Eq.(\ref{telt}) describes, remains to be determined.
The variable $\xi$, as well as $\dot\xi$, rises with time and finally 
the balance of "forces", given by the equation 
\begin{equation}
\label{telt0}
\dot\xi-K^\mu\frac{h_0^{(\theta)}}{h_\mu}\xi^{-\theta/\mu}=0,
\end{equation}
is reached. 
Note that the above expression is equivalent to the Eq.(\ref{telt}) in 
the Markovian limit $\gamma\to\infty$. Therefore 
$p_\gamma(x,t)=p_\infty(x,t)=p(x,t)$ for $t\to\infty$. 
The solution of the Eq.(\ref{telt0}) produces the result 
\begin{equation}
\label{ainf}
a(t)=\left[K^\mu\frac{h_0^{(\theta)}}{h_\mu}
\left(1+\frac{\theta}{\mu}\right)t\right]^{-1/(\mu+\theta)}~~~~~~~(t\to\infty)
\end{equation}
which corresponds to the exact solution (for arbitrary time) for 
the Markovian limit, $p(x,t)$.

The case of the constant diffusion coefficient, $\theta=0$, 
is a particular case and it can be solved easily. The solution of 
the Eq.(\ref{telt}) leads to the following result
\begin{equation}
\label{at0}
a(t)=\frac{1}{K}\left[-\frac{1}{\gamma}(1-e^{-\gamma t})+t\right]^{-1/\mu}.
\end{equation}

For $\theta\ne 0$ and arbitrary time, the Eq.(\ref{telt}) can be
solved by the numerical integration and the distribution $p_\gamma(x,t)$
determined from the Eq.(\ref{solp0}). To evaluate the Fox function 
we use the general formula for its series expansion and then 
the Eq.(\ref{solp0}) can be expressed in the computable form:
\begin{equation}
\label{sze}
p_\gamma(x,t)=\frac{1}{\pi\sigma\mu}\sum_{n=0}^\infty
\frac{\Gamma[1+(2n+1)/\mu]}{(2n+1)!!}(-1)^n\left(\frac{x}{\sigma}\right)^{2n}.
\end{equation}
\begin{figure}[tbp]
\includegraphics[width=8.5cm]{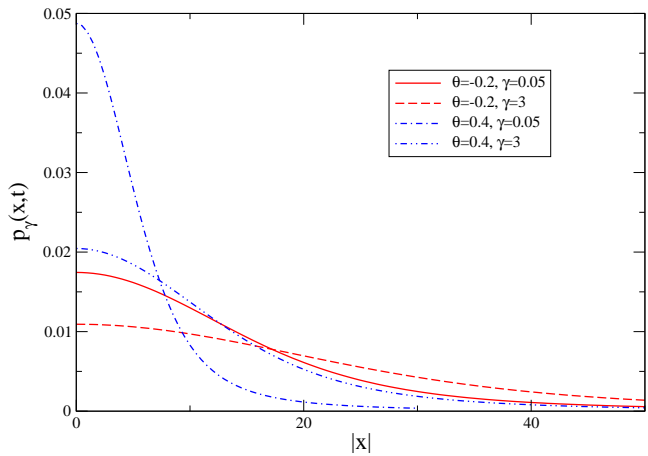}
\caption{(Color online) Probability distributions for the case of the 
exponential kernel with $\mu=1.5$, calculated from the Eqs.(\ref{telt}), 
(\ref{sigmu}), and (\ref{sze}), for $t=50$. The initial condition is 
$p_\gamma(x,0)=\delta(x)$.}
\end{figure}
Fig.1 presents some exemplary probability distributions, so chosen to illustrate 
the influence of the memory on the time evolution. Since the series 
(\ref{sze}) is poorly convergent, the evaluating of the distribution for large 
$|x|$ required the quadruple computer precision \cite{uwa}. 
The case with $\gamma=3$ is close to the Markovian one; 
a comparison with the case characterized by the long memory shows that the spread 
of the distribution slows down with the decreasing value of $\gamma$, 
i.e. for stronger memory (larger "inertia" in the Eq.(\ref{telt})). 
In the limit $t\to\infty$ the curves which correspond to different 
$\gamma$ values and the same $\theta$ -- coincide. 

\section{The general case}

The description by means of the Eq.(\ref{nme}) with the exponential memory kernel 
does not apply to systems with long-time correlations and small decay rate.  
In the study of realistic systems one encounters a variety of forms of the 
kernel; some of them are very complex. It is typical for natural signals that 
they do not represent a simple kinetics, characterised by a unique Hurst exponent. 
Random processes which take into account the whole spectrum of the time-dependent 
Hurst exponents serve then as useful models. This concept, applied to the 
fractional equations formalism, leads to the integration over the order of 
differentiation (the distributed-order diffusion equation) \cite{che} and the 
kernel assumes the integral form: $\int f(\alpha)t^{-\alpha}d\alpha$. Reactions in 
polymer systems are also described by using complicated kernels \cite{cher,ban}.
Therefore, in this Section we consider the Eq.(\ref{nme}) for the case as general as 
possible. We will demonstrate that the solution in a closed form 
can be obtained for the arbitrary kernel.  

The Eq.(\ref{nme}) has the structure of the Volterra integro-differential equation 
with the kernel which depends on the difference of its arguments. Therefore, 
methods using Laplace transforms are applicable. Following Sokolov \cite{sok},  
we apply a method of the integral decomposition which allows us to express 
the required solution by the solution of the corresponding Markovian equation 
(\ref{frace}). Let us define the function $T(\tau,t)$ by its Laplace transform:
\begin{equation}
\label{tdu}
T^\star(\tau,u)=\frac{1}{K_\gamma^\star}\exp\left(-\tau\frac{u}{K_\gamma^\star}\right).
\end{equation}
If we know the function $T(\tau,t)$, the probability distribution 
$p_\gamma(x,t)$ can be obtained by a simple integration:
\begin{equation}
\label{psok}
p_\gamma(x,t)=\int_0^\infty p(x,\tau)T(\tau,t)d\tau.
\end{equation}
However, since the inversion of the Eq.(\ref{tdu}) is difficult for any kernel, 
it is expedient to get rid of $T$. To achieve that, we take the Fourier 
transform from the Eq.(\ref{psok}) and eliminate 
the function $T(\tau,t)$ by using the definition (\ref{tdu}). The final solution is
of the form of the following Fourier-Laplace transform
\begin{equation}
\label{plap}
{\widetilde p}_\gamma^\star(k,u)=\frac{1}{K_\gamma^\star}
{\widetilde p}^\star\left(k,\frac{u}{K_\gamma^\star}\right).
\end{equation}
The above formalism can be applied for any kernel $K_\gamma$ and any operator ${\mathcal L}_x$; 
the main difficulty consists in inversion of the Laplace transforms. 

First of all, we find that if the Markovian process 
$p(x,t)$ is L\'evy distributed in the diffusion limit $k\to 0$, 
then the non-Markovian process is also the L\'evy process in this limit. 
Indeed, the Fourier transform of 
the Markovian solution is given by the Eq.(\ref{pfou}), where $\sigma(t)$ 
follows from the Eqs.(\ref{sigmu}) and (\ref{ainf}). Then we take 
the Laplace transform from that expression and insert the result to the
Eq.(\ref{plap}): ${\widetilde p}_\gamma^\star(k,u)=1/u-F^\star(u)|k|^\mu$.
Finally, the inversion of the Laplace transform yields
\begin{equation}
\label{pfour}
{\widetilde p}_\gamma(k,t)=1-F(t)|k|^\mu,
\end{equation}
which is just the Fourier representation of the L\'evy distribution for small $|k|$. 
To get the function $F(t)$ we need to invert the Laplace transform
\begin{equation}
\label{odwrf}
F^\star(u)=\frac{N h_\mu}{K_\gamma^\star(u)}[a^{-\mu}]^\star(u/K_\gamma^\star)
\end{equation}
and we assume that this inverse transform exists. The solution is given by 
the Eq.(\ref{solp0}), where $\sigma(t)=[F(t)]^{1/\mu}$.

We will consider two particular cases in detail. In the case of the exponential memory 
kernel (\ref{expk}), discussed already in Sec.II, 
we have $K_\gamma^\star(u)=\gamma/(u+\gamma)$ and the Eq.(\ref{plap}) produces 
the following result
\begin{equation}
\label{plap1}
{\widetilde p}_\gamma^\star(k,u)=\frac{1}{u}-a_0\Gamma(1+\alpha)
\gamma^\alpha |k|^\mu u^{-(\alpha+1)}(u+\gamma)^{-\alpha},
\end{equation}
where $a_0=N h_\mu^{1-\alpha}(K^\mu h_0^{(\theta)}\alpha/\mu)^\alpha$
and $\alpha=\mu/(\mu+\theta)$. The above expression cannot be inverted
in closed form. However, if we are interested only in large times, the last term in the
Eq.(\ref{plap1}) can be expanded around $u=0$:
$(u+\gamma)^{-\alpha}\approx\gamma^{-\alpha}-\alpha
\gamma^{-(\alpha+1)}u$. Then the inversion of the Laplace transform
yields
\begin{equation}
\label{pfoure}
{\widetilde p}_\gamma(k,t)=1-a_0|k|^\mu\left(t^\alpha-\frac{\alpha^2}{\gamma}
t^{\alpha-1}\right)~~~~~~~(t\to\infty)
\end{equation}
and this expression demonstrates how the solution $p_\gamma$ approaches its
asymptotic, Markovian form. The final solution, valid for large $t$, is given 
by the Eq.(\ref{solp0}) with $\sigma=[a_0(t^\alpha-\frac{\alpha^2}{\gamma}
t^{\alpha-1})]^{1/\mu}$.

The other physically important kernel has the power-law form, with long tails:
\begin{equation}
\label{potk}
K_\gamma(t)=\frac{t^{-\gamma}}{\Gamma(1-\gamma)}~~~~~(0<\gamma<1).
\end{equation}
The equation (\ref{nme}) with the kernel (\ref{potk}) is usually presented as 
the fractional equation with the Riemann-Liouville derivative
\cite{old} -- which is equivalent to the Caputo operator for a special choice 
of the initial conditions -- in the following form
\begin{equation}
\label{caput}
\frac{\partial p_\gamma(x,t)}{\partial t}= 
{_0}D_t^{\gamma-1}{\mathcal L}_x[p_\gamma(x,t)].
\end{equation}
The power-law kernels are used to describe subdiffusive relaxation e.g. in the framework 
of the CTRW \cite{met}. They emerge also as a result of the coupling 
to the fractal heat bath via the random matrix interaction \cite{lut}.
To solve the Eq.(\ref{nme}) we follow the same procedure as for the exponential kernel. 
The Laplace transform of the Eq.(\ref{potk}) reads $K_\gamma^\star(u)=u^{\gamma-1}$ 
and the Eq.(\ref{plap}) takes the form
\begin{equation}
\label{plap2}
{\widetilde p}_\gamma^\star(k,u)=\frac{1}{u}-a_0\Gamma(1+\alpha)
\gamma^\alpha |k|^\mu u^{-2\alpha+\gamma\alpha-1}.
\end{equation}
The inversion can be easily performed:
\begin{equation}
\label{pfourp}
{\widetilde p}_\gamma(k,t)=1-\frac{a_0\gamma^\alpha\Gamma(1+\alpha)}
{\Gamma(2\alpha-\gamma\alpha+1)} |k|^\mu t^{2\alpha-\gamma\alpha}\equiv 1-F(t)|k|^\mu.
\end{equation}
Clearly, the above solution does not converge with time to the Markovian 
asymptotics, $F(t)\sim t^\alpha$, in contrast to the case of the exponential kernel.

To conclude this Section, we want to mention yet another approach to the 
Eq.(\ref{nme}), which is a direct generalization of the method applied for 
the telegrapher's equation in Sec.II. Inserting the expansion of the functions 
${\widetilde p_\gamma}(k,t)$ and ${\mathcal F}[|x|^{-\theta}p_\gamma(x,t)]$ to 
the Eq.(\ref{nme}) confirms the finding that the solution can be 
expressed in terms of the Fox function $H_{2,2}^{1,1}$ and it is L\'evy 
distributed. The resulting equation is a generalization of the Eq.(\ref{telt}) 
and it determines the function $\sigma(t)$:
\begin{equation}
\label{teltv}
\frac{d\xi}{dt}=K^\mu\frac{h_0^{(\theta)}}{h_\mu}\int_0^t K_\gamma(t-t')\xi^{-\theta/\mu}dt'.
\end{equation}
Mathematically, the Eq.(\ref{teltv}) has the form of the nonlinear Volterra 
integro-differential equation. Since the numerical inversion of the Laplace 
transforms is not always an easy task (methods are often unstable),  
the numerical solving of the Eq.(\ref{teltv}) could be a useful alternative to 
the Eq.(\ref{pfour}).

\section{Diffusion}

The diffusion process is usually characterized by the time dependence of 
the second moment of the probability distribution: if this dependence is linear 
in the limit of long time, the diffusion is called normal. There are many examples of 
physical systems for which the variance rises faster than linearly with time 
(hyperdiffusion), or slower (subdiffusion). Such behaviours are typical for transport 
in the disordered media \cite{bou} and systems with traps and barriers. In the realm 
of dynamical systems, a substantial acceleration of the diffusion is caused by 
the presence of regular structures in the phase space \cite{zas}. On the other hand, 
the subdiffusion appears in the non-Markovian version of CTRW, as a result of 
a non-Poissonian, power-law form of the waiting time distribution \cite{met}. 

When we enter the field of the L\'evy processes, the situation becomes more complicated. 
The stochastic variable performs very long jumps and their size is limited 
only by the size of whole system. As a result, the second moment, 
as well as all moments of the order $\mu$ or higher, is divergent. 
Mathematically, that follows from the fact that the tail of the L\'evy distribution 
is the power-law: $\sim |x|^{-(1+\mu)}$. Therefore, one cannot 
describe the diffusion process in terms of the position variance and  
some other quantity which could serve as an estimation of the speed 
of transport is needed. One possibility is to consider still the second moment
but with the integration limits restricted to a time-dependent interval
(the walker in the imaginary growing box) \cite{jes}. On the other hand,
one can derive the fractional moments of the order $\delta<\mu$.

By the derivation of the moments of the probability distribution $p_\gamma(x,t)$, 
Eq.(\ref{solp0}), we utilize simple properties of the Mellin transform 
from the Fox function:
\begin{equation}
\label{mom}
\langle |x|^\delta\rangle=2\int_0^\infty x^\delta p_\gamma(x,t)dx=
\frac{2}{\mu}\sigma^\delta\chi(-\delta-1)=\frac{2}{\pi}\sigma^\delta
\Gamma(\delta)\Gamma(1-\frac{\delta}{\mu})\sin(\delta\pi/2).
\end{equation} 
Let us consider two quantities: the renormalized moment of order $\mu$, 
defined by the following expression
\begin{equation}
\label{momu}
{\mathcal M}^\mu=\lim_{\epsilon\to0^+}\epsilon\langle |x|^{\mu-\epsilon}\rangle=
\frac{2}{\pi}\sigma^\mu\Gamma(\mu)\sin(\mu\pi/2),
\end{equation}
where we applied the property: $\Gamma(x)\to 1/x$ for $x\to0$, and then
the fractional diffusion coefficient 
${\mathcal D}^{(\mu)}(t)=\frac{1}{\Gamma(1+\mu)}\frac{1}{t}{\mathcal M}^\mu$.
In the Markovian case, defined by the Eq.(\ref{frace}), the
coefficient ${\mathcal D}^{(\mu)}$ is useful to classify
the diffusion: for $\theta<0$ it rises with time, for $\theta>0$ it
falls, and it converges to a constant for $\theta=0$ \cite{sro}. That pattern
is consistent with the diffusion properties, defined in the ordinary sense, of the
Fokker-Planck equation $(\mu=2)$. Therefore, in the following we will name
all kinds of the diffusion -- the subdiffusion, the normal diffusion, and
the superdiffusion -- according to the time dependence of the coefficient 
${\mathcal D}^{(\mu)}$. 

We begin with the case of the
exponential kernel. First we realize that, since $\sigma^\mu=N h_\mu\xi$,
the renormalized moment ${\mathcal M}^\mu$ is directly related to the variable $\xi$: 
${\mathcal M}^\mu=\frac{2}{\pi}N h_\mu\Gamma(\mu)\sin(\mu\pi/2)\xi$. Therefore, the
interpretation of the Eq.(\ref{telt}) is straightforward: it describes
the deterministic time evolution of the moment ${\mathcal M}^\mu$. The diffusion
properties of the system remain unchanged, compared to the Markovian
case, because in the limit $t\to\infty$ both solutions
coincide. However, at small time the influence of the memory, which
hampers both the spread of the distribution and the relaxation to the Markovian 
asymptotics, is visible. Fig.2 illustrates that effect for 
three values of $\theta$ which have different sign. The asymptotic, 
Markovian limit is achieved first for the subdiffusive case 
$\theta=0.4$.
\begin{figure}[tbp]
\includegraphics[width=8.5cm]{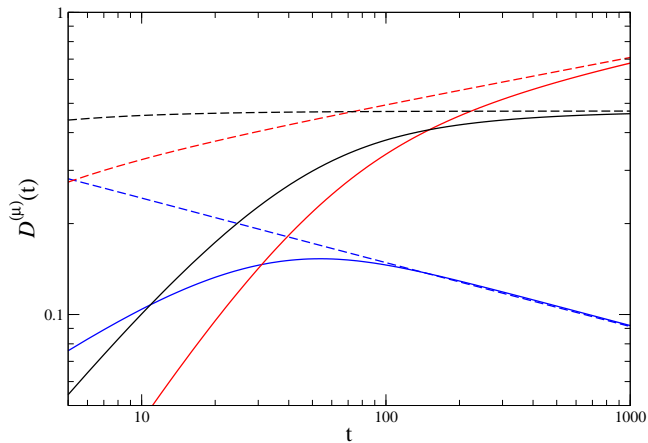}
\caption{(Color online) The fractional diffusion coefficient for the case of the 
exponential kernel with $\gamma=0.05$ (solid lines) and $\gamma=3$ 
(dashed lines), as a function of time, obtained from numerical solving 
of the Eq.(\ref{telt}). Results for three values of
$\theta$ are presented: $\theta=-0.2$ (upper lines for large $t$), 
$\theta=0$ (middle lines), and $\theta=0.4$ (lower lines); $\mu=1.5$.}
\end{figure}

For the case of the power law kernel we calculate the fractional diffusion 
coefficient by means of the Eq.(\ref{momu}); the quantity $\sigma^\mu(t)=F(t)$ 
is given by the Eq.(\ref{pfourp}). We obtain 
\begin{equation}
\label{difp}
{\mathcal D}^{(\mu)}(t)=\frac{2a_0\gamma^\alpha\Gamma(1+\alpha)}
{\pi\mu\Gamma(\alpha-\gamma\alpha+1)}\sin(\mu\pi/2)
t^{2\alpha-\gamma\alpha-1}\sim t^{[\mu(1-\gamma)-\theta]/(\mu+\theta)}.
\end{equation}
The diffusion properties of the system follow directly from the above formula. 
The influence of the parameter $\theta$, which quantifies the structure of the medium, 
is similar as in Markovian case {\cite{sro}: the larger is $\theta$ the weaker is diffusion. 
For $\theta\le0$, there is clearly the superdiffusion. For the positive $\theta$, 
the diffusion becomes weaker with $\theta$ and finally it turns to the 
subdiffusion; the critical value, which corresponds to the normal diffusion, 
is $\theta_{cr}=\mu(1-\gamma)$. On the other hand,
if $0<\theta<\mu$, there is a critical value of $\gamma$ which separates 
the superdiffusion from the subdiffusion: $\gamma_{cr}=1-\theta/\mu$. 
For $\theta>\mu$ the motion is always subdiffusive. The parameter $\gamma$ 
measures the degree of the time nonlocality; it is the largest if
$\gamma$ approaches 0. The diffusion speed grows if $\gamma$ diminishes 
because the system behaviour at large times becomes sensitive to the initial stages 
of the evolution when the distribution spreads rapidly. 
The latter conclusion shows that the memory can influence the diffusion in many ways:
the non-Markovian CTRW predicts the weakening of the diffusion and it is just a consequence 
of the memory in the system \cite{met}. However, in that case the time nonlocality 
invokes a trapping mechanism.  

Note that the above properties, in particular the presence of a transition 
from the subdiffusion to the superdiffusion when changing the parameters of the system, 
still hold if one considers some other fractional moment of order $\delta<\mu$, 
instead of the renormalized moment ${\mathcal M}_\mu$.

For any kernel $K_\gamma$, except for the delta function and for the exponential kernel, 
the time evolution of the moment ${\mathcal M}_\mu$ is governed by the nonlocal 
equation (\ref{teltv}) and the diffusion properties follow from its solution. 
In fact, looking for the full solution may be avoided in some cases: 
the kind of diffusion is already determined by the sign of the function $\ddot\xi(t)$ 
in the limit of long time.

\section{Summary and discussion}

We have studied the diffusion process for the non-Markovian systems with 
the position-dependent diffusion coefficient,  
which involves L\'evy flights and then the variance of the corresponding 
probability distribution is infinite. The integral equation 
for that problem contains the fractional
Riesz-Weyl operator and the time-dependent memory kernel; the diffusion coefficient
depends on the position in the algebraic, scaling way. The equation has been solved 
in terms of the Fox functions in the limit of small wave numbers. We have 
demonstrated that this solution represents 
the L\'evy process for any memory kernel. The formal solution has been obtained 
in the closed form which involves the Laplace transform. The inversion of that transform 
may be a difficult task for the most of the kernels and then numerical methods 
have to be applied. 
Two forms of the kernel have been discussed in details: the exponential kernel, 
for which the problem resolves itself to the generalized telegrapher's equation, 
and power-law one which is equivalent to the fractional 
equation with both the Riesz-Weyl operator and the Riemann-Liouville fractional operator. 
For the exponential kernel, the memory initially slows down 
the spread of the distribution but asymptotically the solution 
converges to that of the Markovian equation. 
The case with the power-law kernel reveals much more interesting behaviour. There is 
an interplay among all ingredients of the dynamics, in particular between the range 
of the memory $\gamma$ and the inhomogeneity parameter $\theta$, which can result in
all kinds of the diffusion, both normal and anomalous.
In order to make that classification possible, 
we have introduced the fractional diffusion coefficient, 
defined in terms of the renormalized moment of order $\mu$. This coefficient allows us 
to maintain the standard terminology of the anomalous diffusion also for the L\'evy flights.


\begin{thebibliography}{99}
\bibitem{uwa}
To determine the tails of the distribution, one can use the expansion in
the negative powers of $|x|$.
\end{thebibliography}
\end{document}